\documentclass[twocolumn,showpacs,preprintnumbers,amsmath,amssymb]{revtex4}

\usepackage{graphicx}
\usepackage{dcolumn}
\usepackage{bm}

\newcommand\ignore[1]{}
\def\one{{\,\hbox{1\kern-.8mm l}}}

\newcommand{\Cset}{{\,\,{{{^{_{\pmb{\mid}}}}\kern-.45em{\mathrm C}}}}}

\newcommand{\cF}{\mathcal F}
\newcommand{\cH}{\mathcal H}

\newcommand{\nn}{\nonumber}

\newcommand{\be}{\begin{equation}}
\newcommand{\ee}{\end{equation}}
\newcommand{\bea}{\begin{eqnarray}}
\newcommand{\eea}{\end{eqnarray}}

\def\a{\alpha}\def\b{\beta}

\def\d{\partial}

\begin{document}

\title{Knotted solutions for linear and nonlinear theories: electromagnetism and fluid dynamics}

\author{Daniel F.W. Alves$^{1}$}\email{dwfalves@ift.unesp.br}
\author{Carlos Hoyos$^{2}$}\email{hoyoscarlos@uniovi.es}
\author{Horatiu Nastase$^{1}$}\email{nastase@ift.unesp.br}
\author{Jacob Sonnenchein$^{3}$}\email{cobi@post.tau.ac.il}
\affiliation{${}^{1}$Instituto de F\'{i}sica Te\'{o}rica, UNESP-Universidade Estadual Paulista, Rua Dr. Bento T. Ferraz 271, Bl. II, 
S\~ao Paulo 01140-070, SP, Brazil}
\affiliation{$^{2}$Department of Physics, Universidad de Oviedo, Calle Federico Garc\'{\i}a Lorca 18, 33007, Oviedo, Spain}
\affiliation{$^{3}$School of Physics and Astronomy, The Raymond and Beverly Sackler Faculty of Exact Sciences,
Tel Aviv University, Ramat Aviv 69978, Israel}
\date{\today}

\begin{abstract}
We examine knotted solutions, the most simple of which is the ``Hopfion'', from the point of view of relations between electromagnetism 
and ideal fluid dynamics. A map between fluid dynamics and electromagnetism works for initial conditions or for linear perturbations, allowing us to 
find new knotted fluid solutions. Knotted solutions are also found to to be solutions of nonlinear generalizations of 
electromagnetism, and of quantum-corrected actions for electromagnetism coupled to other modes. 
For null configurations, electromagnetism can be described as a null pressureless fluid, for which we can  find solutions from the knotted solutions of electromagnetism. 
We also map them to solutions of Euler's equations, obtained from a type of nonrelativistic reduction of the relativistic fluid equations.
\end{abstract}


\maketitle

\section{Introduction}

Solutions with knotted topological structures play an important role in various areas of physics, but in this paper we will concern ourselves 
with two, electromagnetism and fluid dynamics. Despite the simplicity of the theory, the basic knotted solution of free Maxwell electromagnetism 
was only found in 
\cite{Ranada:1989wc,ranada1990knotted}, following an earlier work in \cite{Trautman:1977im}. There are null solutions, $\vec{E}^2-\vec{B}^2=0$, as well as generically non-null solutions in Ra\~nada's construction, both of which are explicitly time dependent. 
In \cite{Kedia:2013bw}, it was shown that 
we can construct more general null solutions obtaining $(m,n)$ knotted structures, and in 
\cite{Hoyos:2015bxa}, new knotted solutions were found using conformal $SO(4,2)$ transformations with complex parameters on known ones. For a review of this subject, and more complete references, see \cite{arrayas2016knots}.

On the other hand, the theory of knots was actually developed in the 19th century based on knotted fluid lines, whose topological robustness 
was already discovered by Lord Kelvin, following the work of Helmholtz in 1858. The abstract study of knots and their evolution \cite{kambe1971motion}
is a fertile subject, for reviews see \cite{ricca1996topological,ricca1998applications,ricca2009new} and the book 
\cite{arnold1999topological}. 
Remarkably, however, explicit theoretical solutions of fluid equations were very scarce, whereas experimental creation of knots 
waited until a few years ago \cite{kleckner2013creation} (see also \cite{proment2012vortex} for some numerical construction). 
Moffat \cite{moffatt1969degree} finally defined a ``helicity'' $\cH_v$ for the fluid flow similar, as we will see, 
to a magnetic helicity for electromagnetism, and wrote some explicit solutions with $\cH_v\neq 0$. The properties of these were studied in 
\cite{moffatt1992helicity,moffatt1992helicity2}. More solutions were found in \cite{ValarMorgulis,crowdy_2004,Ifidon2015216}.
A map of magnetohydrodynamics to just fluid dynamics was used 
extensively (for instance, \cite{Boozer,Kholodenko:2014apa,Kholodenko:2014wfa}), and one also was able to show that 
in magnetohydrodynamics a defined 
``velocity of lines of force'' $\vec{v}_p=(\vec{E}\times \vec{H})/H^2$ \cite{Newcomb} can be measured, and in some cases 
(``frozen field condition'') coincides with the velocity of the fluid transporting it (see for instance \cite{Boozer,Irvine}, the last 
also considering transporting the electromagnetic Hopfion). 

In this letter, we will use connections between electromagnetism and ideal fluid dynamics to find both new knotted solutions in electromagnetism, 
as well as new (time dependent) knotted solutions in fluid dynamics, that we believe have not been written explicitly before.

\section{Knots in electromagnetism}

In this section we review electromagnetic knotted solutions and some of their properties.
Using the Riemann-Sielberstein (RS) vector $\vec{F}=\vec{E}+i\vec{B}$ ($c=1$) , the source-free Maxwell's equations are
\be\label{maxwell}
\vec{\nabla}\times \vec{F}=i\frac{\d}{\d t}\vec{F};\;\;\; \vec{\nabla}\cdot \vec{F}=0\;,
\ee
Electromagnetic duality is manifest as a rotation $\vec{F}\to e^{i\phi}\vec{F}$. We will define electric and magnetic potentials as $\vec{F}=\vec{B}_e+i\vec{B}_m=\vec{\nabla}\times (\vec{A}_e+i\vec{A}_m)$. To characterize the non-trivial topology of electromagnetic fields, a common set of observables are the helicities, that give a measure of the mean value of the linking number of the electromagnetic field lines. The helicities $\cH_{ab}$, $a,b=e,m$ are defined as
\be
\cH_{ab}=\int d^3x \vec{A}_a\cdot \vec{B}_b\;.
\ee
Calculating their time derivatives, we find that $\cH_{ee}$ and $\cH_{mm}$ are conserved if $\vec{E}\cdot \vec{B}=0$ 
and $\cH_{em}$ and $\cH_{me}$ are conserved if $\vec{E}^2-\vec{B}^2=0$.  One can find solutions that have conserved helicites and contain ``knotted'' structures for the electric and magnetic fields, characterized by Hopf or winding number invariants of the field structures. 

\subsection{Knotted solutions in Bateman's construction} 

One way to obtain knotted solutions is in Bateman's construction, imposing an ansatz
\be\label{bateman}
\vec{F}=\vec{\nabla}\a \times \vec{\nabla}\b\;,
\ee
with $\a,\b\in \mathbb{C}$. One Maxwell equation, $\vec{\nabla}\cdot \vec{F}=0$, is automatically satisfied, and the other imposes the 
constraint $\vec{F}^2=0$ $\Rightarrow$ $\vec{E}^2-\vec{B}^2=0$ and $\vec{E}\cdot\vec{B}=0$, so these are {\em null} solutions. Using this construction one can show that there is a ``Hopfion'' solution
\be\label{hopfsol}
\a=\frac{r^2-t^2-1+2iz}{r^2-t^2+1+2it}; \ \ \b=\frac{2(x-iy)}{r^2-t^2+1+2it}.
\ee
Where $(t,x,y,z)$ are the spacetime coordinates and $r^2=x^2+y^2+z^2$. The Hopfion has $\cH_{em}=\cH_{me}=0$ and nonzero helicities $\cH_{mm}=\cH_{ee}$, but their value depends on the amplitude of the electromagnetic field $|\vec{E}|$. An invariant that only depends on the topology can be constructed by introducing a map to unit quaternions $\mathfrak{q}=(\a+\b j)/\sqrt{|\alpha|^2+|\beta|^2}\in \mathbb{H}$ \cite{Hoyos:2015bxa}. Using $\mathbb{R}^3\cup\{\infty\}\cong S^3$ and $\mathbb{H}_{|\mathfrak{q}|^2=1}\cong SU(2)\cong S^3$, at any fixed time the ``Hopfion'' solution maps $S^3\rightarrow S^3$ with unit winding number $w=1$. By replacing $\a$ with $\a^m$ and $\b$ with $\b^n$ in \eqref{bateman}, we find a $(m,n)$ knot solution with winding number $w=mn$. 

\subsection{Knotted solutions in Ra\~nada's construction}

There are more general, {\em non-null}  ($\vec{E}^2-\vec{B}^2\neq 0$), knotted solutions. 
Consider the ansatz for $F_{\mu\nu}$ and $*F_{\mu\nu}$ of the type
\bea
F_{\mu\nu}&=& \frac{\sqrt{a}}{2\pi i}\frac{1}{(1+\bar\phi\phi)^2}(\d_\mu\bar\phi\d_\nu\phi-\d_\nu\bar\phi\d_\mu\phi)\;,\cr
*F_{\mu\nu}&=& \frac{\sqrt{a}}{2\pi i}\frac{1}{(1+\bar\theta\theta)^2}(\d_\mu\bar\theta\d_\nu\theta-\d_\nu\bar\theta\d_\mu\theta). 
\label{ranada}
\eea
The condition $*F_{\mu\nu}=\frac{1}{2}\epsilon_{\mu\nu\rho\sigma}F^{\rho\sigma}$, determines the equations for $\phi$,$\theta$. The solutions then solve Maxwell's equations, and by construction $F_{\mu\nu}*F^{\mu\nu}\propto \vec{E}\cdot\vec{B}=0$. Moreover, the 2-form $F$ decomposes as $F=dq\wedge dp$ (Clebsch representation), where $p,q$ are {\em real} functions, and in particular
\be
\vec{B}=\vec{\nabla}p\times \vec{\nabla}q\;.
\ee
The relation between $p,q$ and $\phi$ is 
\be\label{pqvals}
p=\frac{1}{1+|\phi|^2};\;\;\; q=\frac{arg(\phi)}{2\pi}\;,
\ee
and there is a similar one for $*F=du\wedge dv$, with $u,v$ in terms of $\theta$.
If $p$ and $q$ are {\em single-valued and well-defined in the whole of space}, then the magnetic 
helicity $\cH_{mm}$ is zero. If not, we have
\be
\vec{A}=p\vec{\nabla}q+\vec{\nabla}\chi\;,\label{vecAdecomp}
\ee
where $\chi$ is such that $\vec{A}$ is well defined. The magnetic helicity is 
\be
\cH_{mm}=\int d^3x \vec{\nabla}\chi\cdot(\vec{\nabla} p\times \vec{\nabla}q).
\ee
Defining the function
\be
\phi_H(x,y,z)=\frac{2(x+iz)}{2z+i(r^2-1)},\label{phiH}
\ee
the ``Hopfion'' solution in Ra\~nada's construction is such that, at $t=0$,
\be\label{hopfranada}
\phi(x,y,z)=\phi_H(z,-y,x),\ \ \theta(x,y,z)=\phi_H(x,z,-y).
\ee
The solution has $\cH_{ee}=\cH_{mm}\neq 0$ and $\cH_{em}=\cH_{me}=0$. A more general set of solutions with non-zero helicities \cite{Arrayas:2011ia} are, at $t=0$, 
\be\label{knotsranada}
\phi= \frac{(x+iy)^{(n)}}{(z+i(r^2-1)/2)^{(m)}}\ \ \theta=\frac{(y+iz)^{(l)}}{(z+i(r^2-1)/2)^{(s)}}\;.
\ee
where the $(m)$ index means we leave the modulus intact, but we raise the phase to the $m$-th power.

\subsection{Hopf index}

Using $\mathbb{R}^3\cup\{\infty\}\cong S^3$ and $\mathbb{C}\cup\{\infty\}\cong S^2$, at any fixed time $\phi$ can be seen as a map $S^3\rightarrow S^2$, and the magnetic helicity is its Hopf index $\cH(\phi)$.  Indeed, we can define an area 2-form on $S^2$,
\be
{\cal \omega}=\frac{1}{2\pi i}\frac{d\phi^*\wedge d\phi}{(1+|\phi|^2)^2}\;,
\ee
whose pullback onto $S^3$ is a 2-form $\cF$ whose components are the spatial components in \eqref{ranada} (taking $a=1/4$) $\cF_{ij}=F_{ij}$. The Hopf index then equals the magnetic helicity
\be
\cH(\phi)=\int_{S^3}{\cal A}\wedge {\cal F}=\int d^3 x\, \vec{A}\cdot\vec{B}=\cH_{mm}\;.
\ee
A Hopf index $\cH(\phi_H)=1$ is found for instance for \eqref{phiH}.

\section{Solutions to nonlinear theories}

We now show that the knotted solutions in Bateman's construction are also solutions of any nonlinear electromagnetism theory that reduces to the Maxwell case for small fields. This is true in general for any null configurations (for recent disscusions see e.g. \cite{Ortaggio:2015rra,Goulart:2016orx}).
Since $\vec{F}^2=0$, both the two possible Lorentz invariants constructed out of $\vec{E}$ and $\vec{B}$ vanish on the solutions.  Let us define
\bea
L&\equiv& \frac{F_{\mu\nu} F^{\mu\nu}}{2b^2}=\frac{1}{b^2}(\vec{B}^2-\vec{E}^2);\cr
P&\equiv&
\frac{1}{8b^2}\epsilon^{\mu\nu\rho\sigma}F_{\mu\nu}F_{\rho\sigma}= \frac{1}{b^2}\vec{E}\cdot \vec{B}\;,
\eea
where $b$ is a dimension 2 constant. We assume that fields vary on distances much larger than $b^{-1/2}$, so that we can ignore possible derivatives on $F_{\mu\nu}$. Then, nonlinear generalizations of electromagnetism are described by actions of the form
\be
{\cal L}=b^2\left[-\frac{L}{2}+\sum_{n\geq 2}\sum_{m\geq 0}c_{n,m}L^nP^m\right]\;,\label{expan}
\ee
i.e., which reduces to Maxwell at small fields, and has nonlinear corrections written solely in terms of $L$ and $P$. 
This includes actions like the Born-Infeld Lagrangian, ${\cal L}=-b^2[\sqrt{1+L-P^2}-1]$ \cite{Born:1934gh},
introduced to get rid of diverging electric fields in electromagnetism, as well as actions obtained by integrating out other 
fields, like in the case of the one-loop Euler-Heisenberg Lagrangian for QED, where the fermions have been integrated out,  
${\cal L}=b^2\left[-\frac{L}{2}+\b b^2[L^2+7P^2]\right]$, where $b=m^2$ and $\b =2\a^2/45$ ($m$ is the fermion mass and $\alpha$ 
the fine structure constant). This form also applies for {\em any higher loop} integration of the coupling to {\em any field}
(see \cite{Dunne:2004nc} for a review). Therefore knotted solutions are valid in the full 
quantum theory after integrating out the other fields.  Moreover, this is not only true for usual quantum fields, but also string modes
in string theory can be integrated out to give the same result. 
Indeed, for electromagnetism confined to a D-brane in string theory, the integrating out of higher modes (in $\a'=l_s^2$, the string scale)
results in the BI action with $b=1/\a'$. Note that in this case, the action is $\a'$ exact at leading order in $g_s$, and there are no $\d_\mu F_{\nu\rho}$ 
terms.

In terms of quantities analogous to the ones of electromagnetism in a medium (see \cite{Nastase:2015ixa}),
\be
\vec{H}\equiv -\frac{\d {\cal L}}{\d \vec{B}},\;\;\;\;
\vec{D}\equiv +\frac{\d {\cal L}}{\d \vec{E}}\;,
\ee
the equations of motion look formally the same as Maxwell's in a medium, 
\bea
\vec{\nabla}\times \vec{E}+\d_t \vec{B}=0, && \vec{\nabla}\cdot \vec{B}=0\cr
\vec{\nabla}\times\vec{H}-\d_t\vec{D}=0, && \vec{\nabla}\cdot \vec{D}=0\;.
\eea
For solutions such that $L=P=0$, $\vec{H}$ reduces to $\vec{B}$ and $\vec{D}$ to $\vec{E}$, so we have Maxwell's equations in vacuum and thus knotted solutions found using Bateman's construction in Maxwell theory are solutions of the nonlinear theory as well.

\section{Mapping electromagnetic to fluid knots}

An ideal fluid, for an adiabatic flow with potential per unit mass $\pi$ is governed by the Euler's and continuity equations, 
\bea
&&\d_t\vec{v}+(\vec{v}\cdot\vec{\nabla})\vec{v}=-\frac{1}{\rho}\vec{\nabla}p-\vec{\nabla}\pi\label{Euler}\\
&&\d_t\rho+\vec{\nabla}\cdot (\rho \vec{v})=0.\nn
\eea
The right hand side of the Euler's equations (\ref{Euler}) equals $-\vec{\nabla}h$ ($\delta h =\frac{\delta p}{\rho}+\delta \pi$), 
where $h$ is the enthalpy per unit mass. The continuity equation can also be rewritten as 
\be
\d_t h+\vec{v}\cdot\vec{\nabla}h+c_s^2\vec{\nabla}\cdot\vec{v}=0\;,\label{cont}
\ee
where $c_s=\sqrt{\d p/\d \rho}$ is the sound speed. 

For incompressible fluids $\vec{\nabla}\cdot\vec{v}=0$ and compressible barotropic fluids $p=p(\rho)$, there is also a conserved helicity. From the Euler's equations (\ref{Euler}), one finds 
\be
\d_t(\vec{v}\cdot\vec{\omega})-\vec{\nabla}\cdot\left[\vec{\omega}\left(\frac{\vec{v}^2}{2}-\int \frac{dp}{\rho}-\pi\right)-\vec{v} 
(\vec{v}\cdot\vec{\omega})\right]=0\;,
\ee
where $\vec{\omega}=\vec{\nabla}\times\vec{v}$ is the vorticity,
which means that we have the conserved {\em fluid helicity}, the integral of the velocity Chern-Simons term,
\be
\cH_v=\int d^3x \,\vec{v}\cdot \vec{\omega}=\int d^3x\, \vec{v}\cdot(\vec{\nabla}\times\vec{v})\,.
\label{fluidhel}
\ee
Knotted fluid solutions are solutions for which there is a linking of $\vec{v}(t,\vec{x})$ at fixed time $t$, i.e. nonzero fluid helicity. 

We see the analogy with electromagnetism \cite{Marnanis:1998,Sridhar:1998} : $\vec{v}$ is the analog of $\vec{A}$, so $\vec{B}$ is the analog of the vorticity $\vec{\omega}$, 
and $\cH_v$ is the analog of $\cH_{mm}$. One can define Clebsch variables $\lambda$ and $\mu$ for the fluid, as in \cite{kuznetsov1980topological}, 
giving the velocity field
\be
\vec{v}=\lambda\vec{\nabla}\mu +\vec{\nabla}\Phi\;,\label{velclebsch}
\ee
where $\Phi$ is the fluid potential, just like the decomposition for $\vec{A}$ in (\ref{vecAdecomp})
in the Ra\~nada construction.
Note that this is not the much-used map from magneto-hydrodynamics (fluid coupled to electromagnetism) to hydrodynamics, 
where $\vec{v}$ is mapped to $\vec{v}$, but $\vec{\omega}$ is mapped to $\vec{B}$, and one restricts the configurations to the 
ones with $\vec{B}=\vec{\omega}=\vec{\nabla}\times \vec{v}$. Instead, we can define a full map, from fluid to electromagnetism, by 
\be
\vec{B}=\vec{\omega}=\vec{\nabla}\times\vec{v},\;\;\;\;
\vec{E}=-\d_t \vec{v}-\vec{\nabla}h\;,
\ee
which means that really $\vec{A}=\vec{v};\;\;\; A_t=h$, though we have no gauge invariance now, since $\vec{v}$ is physical (observable).  Two of Maxwell's equations  $\vec{\nabla}\cdot\vec{B}=\vec{\nabla}\times\vec{E}+\partial_t\vec{B}=0$ are automatically satisfied, but from the Euler's equations (\ref{Euler})
we find the condition
\be
\vec{E}=(\vec{v}\cdot\vec{\nabla})\vec{v}=\vec{\omega}\times \vec{v}+\vec{\nabla}\left(\frac{\vec{v}^2}{2}\right)=
\vec{B}\times \vec{A}+\vec{\nabla}\left(\frac{\vec{A}^2}{2}\right)\;,
\ee
which does not hold for knotted solutions of Maxwell's equations. At the linearized level however, it implies $\vec{E}=0$, i.e., pure magnetism, and the continuity equation, in electromagnetic variables
\be
\d_t A_t+(\vec{A}\cdot \vec{\nabla})A_t+c_s^2\vec{\nabla}\cdot \vec{A}=0\;,\label{lorentz}
\ee
becomes the Lorenz gauge condition, identifying $c_s$ with the speed of light.

Rather than mapping the full solution, we will use the map between electromagnetism and fluid variables at a fixed time
\be
\vec{v}(t=0,\vec{x})=\vec{A}(t=0,\vec{x}), \ h(t=0,\vec{x})=h_0.
\ee
Where $h_0$ is an arbitrary constant. In this case the knotted solutions supply initial conditions for Euler's and continuity equations. The time-dependent solutions have non-zero helicity and correspond to a fluid configurations of non-trivial topology.

In \cite{kuznetsov1980topological}, knotted solutions for an incompressible fluid were (implicitly) found by giving an initial condition of the form \eqref{velclebsch}, where $\lambda=\cos\vartheta(\vec{x}),\mu=\varphi(\vec{x})$, with $(\vartheta,\varphi)$ the polar and azimuthal angles of a $S^2$.
$\Phi$ would be determined by the incompressibility condition, but finding an explicit solution is the main obstacle to obtaining an analytic expression for the velocity. 
We will avoid this issue by considering more general cases of compressible barotropic fluids, so the only constraint on $\Phi$ is that the velocity should be a smooth function of the spatial coordinates. Knotted solutions in Ra\~nada's construction can be mapped for instance using the Clebsch decompositions \eqref{vecAdecomp} and \eqref{velclebsch}, making $\lambda=p$, $\mu=q$ and $\Phi=\chi$, where $p,q$ are given by \eqref{pqvals} and $\phi$ can be taken to be \eqref{hopfranada} or \eqref{knotsranada}. In terms of $\phi$, the velocity is
\be
\vec{v}=\frac{1}{4\pi i (1+|\phi|^2)}\left(\frac{\vec{\nabla}\phi}{\phi}-\frac{\vec{\nabla}\phi^*}{\phi^*} \right)+\vec{\nabla}\Phi\;.
\ee
In addition, we present here an additional set of initial conditions for knotted solutions. Consider the stereographic projection of $S^3$ on $\mathbb{R}^3$:  
\be
X_i=\frac{2 x_i}{1+r^2}, \ \ X_4=\frac{1-r^2}{1+r^2}.
\ee
We define the complex coordinates $ Z_1=X_1+i X_2$,  $Z_2=X_3+i X_4$, that then we use them to define a $\phi$
\be
\phi =\frac{Z_1^n}{Z_2^m}, \ n,m\geq 1 ,\ \ n,m\in \mathbb{Z}\;.
\ee
A non-singular velocity is obtained for
\be
\Phi=-\frac{n}{2\pi} \arctan \left(\frac{y}{x}\right).
\ee
The helicity of these configurations is $\cH_v=-nm$. 

\section{Electromagnetism as a fluid and its knotted solutions}

Any gapless quantum system is expected to have an effective fluid description at low energies. For a relativistic system  this means that the energy-momentum tensor can be put in the form
\be
T_{\mu\nu}=\rho u_\mu u_\nu +p(\eta_{\mu\nu}+u_\mu u_\nu)+\pi_{\mu\nu}\;,
\ee
where $u_\mu$ is the $4$-velocity of the fluid, $\rho$ is the energy density and $p$ the pressure. $\pi_{\mu\nu}$ depends on derivatives of $u_\mu$, $\rho$ and $p$. From $\d_\mu T^{\mu\nu}=0$ we obtain the relativistic fluid equations.
This program can be applied to any quantum system 
as in \cite{Nastase:2015ljb,Endlich:2010hf,Dubovsky:2011sj,Berezhiani:2016dne}. For a perfect fluid $\pi_{\mu\nu}=0$ we obtain in the non-relativistic limit
$u^\mu\simeq (1,\vec{v})$, $|\vec{v}|^2\ll 1$, $p\ll  \rho$ the continuity and Euler's equations. However, in electromagnetism we can also find another, more unusual, fluid, 
a ``null pressureless fluid'', or ``null dust'', with $p=0$ and $u^\mu u_\mu=0$. 

As has been observed in \cite{Birula:1992}, this is the case for null configurations with $\vec{E}^2-\vec{B}^2=0$ and $\vec{E}\cdot\vec{B}=0$. In this case the electromagnetic energy-momentum tensor takes the form $T_{\mu\nu}=\rho u_\mu u_\nu$, with  $u_\mu=(1,\vec{v})$  ($\vec{v}^2=1$) and
\be
\rho= \frac{1}{2}(\vec{E}^2+\vec{B}^2);\;\;\;\;
\vec{v}=\frac{1}{\rho}\vec{E}\times \vec{B}\;.\label{velmap}
\ee

\begin{figure}[t!]
\begin{tabular} {c}
\includegraphics[width=6cm]{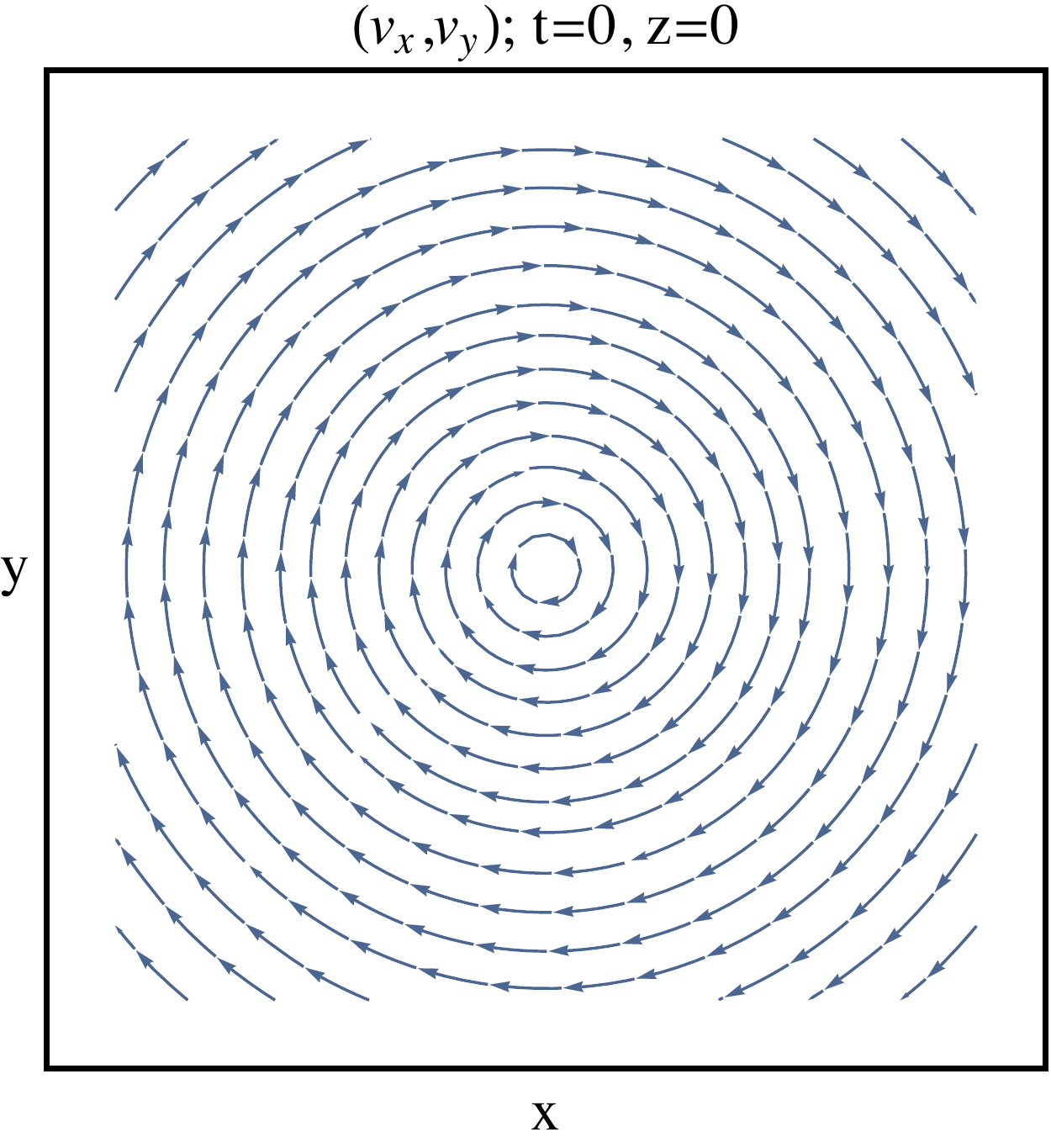} \\ \includegraphics[width=6cm]{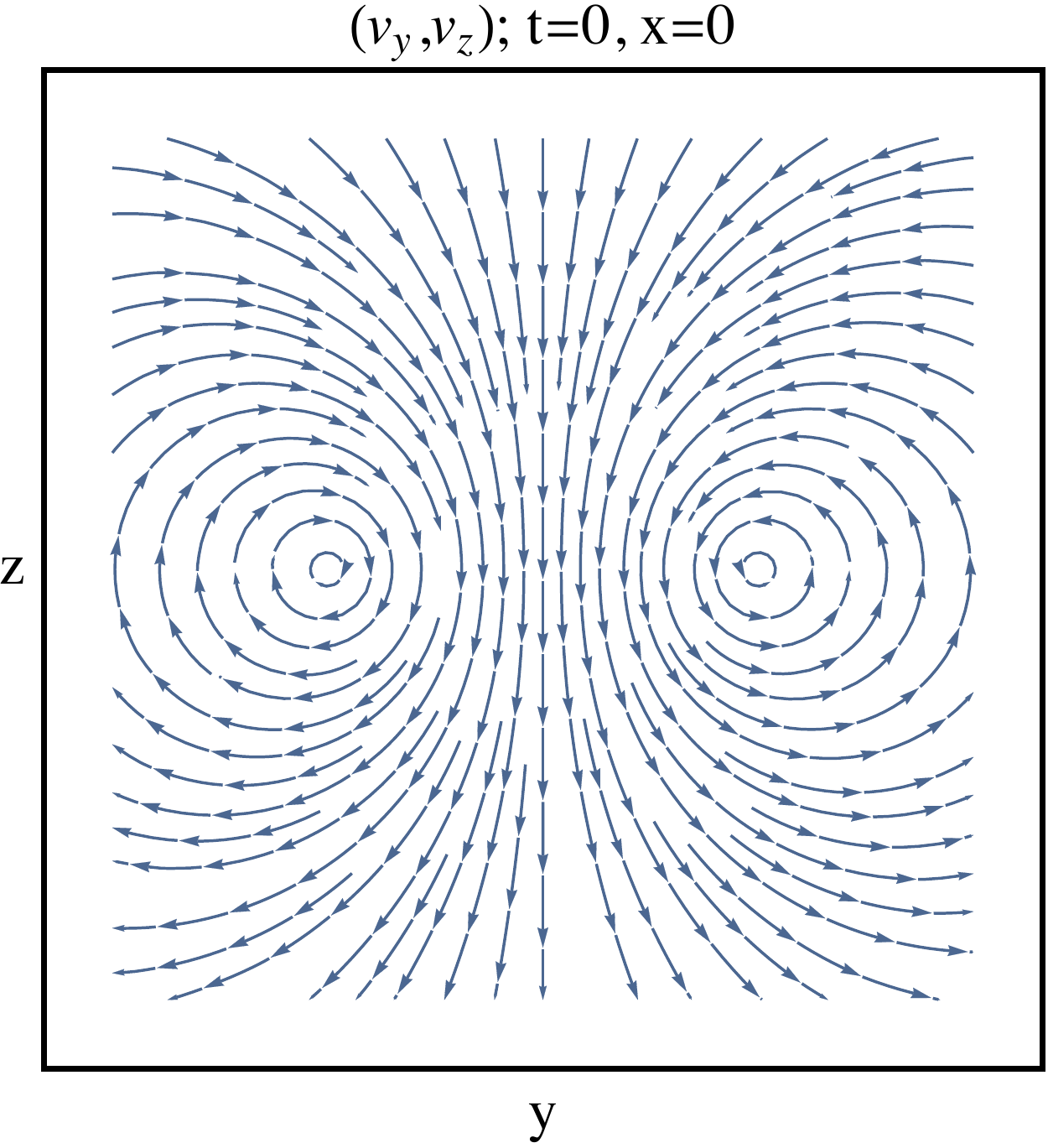}
\end{tabular}
\caption{Orthogonal sections of the velocity field for the Hopfion solution, on the $(x,y)$ plane (top) and $(y,z)$ plane (bottom).  Using rotational symmetry in the $(x,y)$ directions the linked torus structure is apparent.}
\label{fighopf}
\end{figure}

The Hopfion solution determined by \eqref{bateman} and \eqref{hopfsol} is null, so we can map it to fluid dynamics. We find an energy density
\be\label{rhohopf}
\rho = \frac{16 \left((t-z)^2+x^2+y^2+1\right)^2}{\left(\left(r^2-t^2+1\right)^2+4t^2\right)^3}.
\ee
The velocity field is
\be\label{velhopf}
v_x=\frac{2(y+x(t-z))}{1+x^2+y^2+(t-z)^2},\;
v_y=\frac{-2(x-y(t-z))}{1+x^2+y^2+(t-z)^2},
\ee
and $v_z^2=1-v_x^2-v_y^2$. The topological structure can be seen in Fig.~\ref{fighopf}. Although the fluid helicity ${\cal H}_v$ diverges due to non-vanishing contributions at spatial infinity, these can be subtracted to produce a finite nonzero result. The velocity of the null fluid can be rewritten in terms of the RS vector $\vec{F}$ and its complex conjugate $\vec{F}^*$, as 
\begin{equation}
\vec{v}=2i\vec{F}\times \vec{F}^*/\vec{F}\cdot \vec{F}^*.\label{velF}
\end{equation}
This implies that $(m,n)$ knots, and more general null solutions found by the replacement in \eqref{bateman}
$\alpha\to f(\alpha)$, $\beta\to g(\beta)$, have the same velocity as the Hopfion. On the other hand, the energy density $\rho$ is different for each configuration.

\subsection{Maps to non-relativistic fluids}

Switching to lightcone coordinates, $x^\pm =t\pm z$, the velocity of the Hopfion solution (\ref{velhopf}), is independent of $x^+$, in such a way that it can be mapped to a solution of a non-relativistic $2+1$-dimensional system, with $\tau=x^-$ playing the role of time coordinate. Defining $\b^a$ by $\beta^a=\frac{v^a}{1-v^z}$, $a=x,y$, for any configuration satisfying $\d_+ v^a=0$, the relativistic fluid equations become the $2+1$-dimensional Euler's equations of a compressible fluid with velocity $\beta^a$ and constant pressure \footnote{This map should not be confused with other non-relativistic limits of electromagnetism where the speed of light is taken to infinity \cite{Rousseaux:2013}.}:
\be
\partial_\tau\beta^a+ \beta^b\partial_b\beta^a=0.\label{betabeta}
\ee
In this description, the Hopfion solution is a spherical bouncing shock 
\be
\beta^a=\epsilon^{ab}\partial_b \tilde{\psi}+\tau\partial^a\tilde{\psi},\ \ \tilde{\psi}=\log(x^2+y^2-1-\tau^2).
\ee

If $\d_+ v^a=0$, it is also possible to establish a map to an {\em incompressible} fluid. The relativistic fluid equations are
\be
(1-v^z)\partial_\tau v^a+v^b\partial_b v^a=0.
\ee
At any fixed $\tau$ we can make an identification of the first term with the pressure in Euler's equations
\be
\partial^a p\equiv (1-v^z)\partial_\tau v^a\;.\label{Pmap}
\ee
Then $v^a$, $p$ map to {\em steady state} (time-independent) 
solutions of Euler's equations, with $\tau$ seen as a parameter. The Hopfion solution can be written as 
\be
v^a=\epsilon^{ab}\partial_b \psi+\tau \partial^a\psi,\ \ \psi=\log(1+x^2+y^2+\tau^2)\;.
\ee
At $\tau=0$ the velocity satisfies the incompressibility condition $\partial_a v^a=0$.
In this case, an explicit solution to \eqref{Pmap} can be found  
\be
p=p_\infty-\frac{2}{1+x^2+y^2}.
\ee
This solution is a smooth vortex, it can be obtained by the stereographic projection of a constant vorticity configuration on the sphere (see e.g.\cite{crowdy_2004}). 
This map can also be used to give initial conditions, at $\tau=0$, for the null fluid, given a steady state ($v_S^a, p_S$) solution in 2+1 dimensions, 
by 
\be
v^a(\tau=0)=v_S^a,\ \ \partial_\tau v^a(\tau=0)=\frac{\partial^a p_S}{1-v_S^2}.
\ee

\section{Acknowledgements}
We would like to thank Manuel Array\'as, Ori Ganor and Nilanjan Sircar for useful comments and discussions.
This work was supported in part by a center of excellence supported by the Israel Science Foundation (grant number 1989/14), and by 
the US-Israel bi-national fund (BSF) grant number 2012383 and the Germany Israel bi-national fund GIF grant number I-244-303.7-2013. 
The work of HN is supported in part by CNPq grant 304006/2016-5 and FAPESP grant 2014/18634-9. HN would also like to 
thank the ICTP-SAIFR for their support through FAPESP grant 2016/01343-7. C.H. is supported by the Ramon y
Cajal fellowship RYC-2012-10370, the Asturian grant FC-15-GRUPIN14-108 and the Spanish national
grant MINECO-16-FPA2015-63667-P. D.F.W.A. is supported by CNPq grant 146086/2015-5.

\end{document}